\documentclass[a4paper]{jpconf}

\usepackage{graphicx}

\begin{document}

\title{Merger of Massive Black Holes using {\it N}-Body Simulations with Post-Newtonian Corrections}

\author{Miguel Preto$^1$, Ingo Berentzen$^1$, Peter Berczik$^{1,2}$, David Merritt$^3$ and Rainer Spurzem$^1$}...

\address{$^1$ Astronomisches Rechen-Institut, Zentrum f\"{u}r Astronomie,
         M\"{o}nchhofstr. 12-14, 69120 Heidelberg, Germany}

\address{$^2$ Main Astronomical Observatory, National Academy of Sciences of
  Ukraine, 27 Akademika Zabolotnoho St., 03680 Kyiv, Ukraine}

\address{$^3$ Center for Computational Relativity and Gravitation, Rochester
  Institute of Technology, 78 Lomb  Memorial Drive, Rochester, NY 14623, USA}

\ead{miguelp@ari.uni-heidelberg.de}

\begin{abstract}

We present preliminary results from self-consistent, high resolution direct
{\it N}-body simulations of massive black hole binaries in mergers of galactic
nuclei. The dynamics of the black hole binary includes the full Post-Newtonian
corrections (up to 2.5PN) to its equations of motion. We show that massive black
holes starting at separations of $100$ pc can evolve down to
gravitational-wave-induced coalescence in less than a Hubble time. The
binaries, in our models, often form with very high eccentricity and, as a
result, reach separations of $50$ Schwarzschild radius with eccentricities which are 
clearly distinct from zero --- even though gravitational wave emission damps
the eccentricity during the inspiral. These deviations from exact circular 
orbits, at such small separations, may have important consequences for LISA
data analysis.

\end{abstract}

\section{Introduction}

Massive black holes (henceforth MBHs) are ubiquituous in the centers of
galaxies. During the last decade, a great deal of effort was made
observationally to characterize the physical properties of galactic nuclei
and their black holes \cite{FF05}. A number of phenomenological relations 
were unveiled: the masses of the central MBH appear to be correlated with 
some physical properties of the host nucleus, namely the luminosity
\cite{Ma98} and the velocity dispersion \cite{FM00,Ge00}. Massive galaxies, 
however, typically do not evolve in isolation and undergo several merger
events during their lifetime. If each of the merging galaxies harbors a MBH 
at the center, then a MBH binary is expected to form in the core of the
resulting remnant. 

The paradigm for MBH binary evolution consists essentially of three distinct
phases \cite{BBR80}: (i) Right after the merger the two MBHs sink on a
dynamical friction time scale,  towards the center of the remnant where they 
form a bound pair; (ii) Once such bound pair is formed, its semimajor axis 
continues to shrink by dynamical friction until the binary becomes {\it hard}; 
afterwards, the binary continues to harden by the
ejection of field stars; (iii) When the binary's separation becomes small 
enough, gravitational wave (henceforth GW) emission may drive the final stage
of relativistic inspiral and coalescence. The transition from (i) to (ii) is
understood to be unproblematic, as it typically occurs on a cosmologically short
time scale ($\leq 10^{8-9}$ yr.) as long as the mass ratio is $q = M_1/M_2 \geq 0.1$ \cite{MRev06};
the subsequent transition from Phase (ii) to Phase (iii), however, could constitute
a bottleneck for the evolution towards relativistic inspiral and coalescence. 
The binary stalling problem is especially severe in the case of both
purely stellar (ie. with no gas), spherically symmetric galaxy models. In this
case, the pool of stars available for close interaction with the MBH binary is 
very rapidly exhausted (roughly on the order of a local dynamical time scale),
and leaves the binary dynamically isolated at separations of the order
of a parsec, corresponding roughly to the ``hard binary separation'' 
$a_h \equiv G \mu_{red}/4 \sigma^2$, where $\mu_{red}$ is the binary's reduced
mass and $\sigma$ is the velocity dispersion of the surrounding stellar
field. This is the so-called {\it Final Parsec Problem} (hereafter
FPP). If the time scale on which stars can diffuse in angular momentum space to
enter the binary's loss cone is longer than a Hubble time, then the MBH binary 
evolution will stall and it will not become a GW source detectable by
LISA. There are many proposed 
solutions to the FPP: dissipation of the binary's orbital energy in a gaseous 
circumbinary disk \cite{AN05,Mayer07},  massive perturbers \cite{PA08}, the
infall of a third (Intermediate-)MBH, a triaxial potential may keep the loss cone full 
\cite{MPoon04,BMSB06,BP08}, etc. 

Here we present preliminary results of {\it N}-body simulations of idealized models
of merging spherical galactic nuclei, starting from an initial separation of 
tens of parsec. We follow the subsequent formation and evolution of the MBH binary 
until it reaches the final stages of relativistic inspiral at separations of a
few Schwarschild radii ($\sim$ few $\times 10^{-7}$ pc, if the MBHs have masses
$\sim 10^6 M_\odot$). The nucleus that
results from the merger has a pronounced triaxial structure; triaxiality 
facilitates the MBH binary's transition to the GW regime, presumably by
inducing a collisionless regime which is capable of repopulating the loss cone 
due to centrophilic orbits. In sec. 2, we 
describe the galaxy merger set-up and the formation of the bound MBH pair.
In sec. 3, we show that the MBH binary reaches the GW-dominated inspiral 
phase while the merger remnant still retains some degree of triaxiality.
In sec 4, we show that the MBH binary may reach the LISA band with an 
eccentricity different from zero. We conclude with Sec. 5.

\section{Simulation of the merger of two spherical galactic nuclei}

A number of studies were dedicated to the FPP for MBH binaries ---
{\it placed} in the center of a spherical model of a galaxy merger remnant ---
using the Fokker-Planck formalism, {\it N}-body simulations, or both
\cite{Yu02,MM03,Hem02,MF04,BMS05,MMS07}. 
In those models, the loss cone is mostly empty and repopulation of loss cone
orbits is driven by two-body relaxation in the diffusive regime \cite{CK78,MM03}. The 
characteristic time scale 
for this process is the {\it relaxation time}, which is proportional
to the total number $N$ of stars in the nucleus. For the number of stars present in a 
typical galactic nucleus, this generally exceeds the Hubble time by a large 
margin --- the
exception being the most compact and dense nuclei. From cosmological studies,
we expect that the mass function of MBH binaries peaks in the
$10^5-10^6 M_\odot$ range and at high redshift $z \geq 4$ \cite{Sesana07}; in 
such cases, the diffusive refilling of the loss cone may be marginally
efficient, even in the spherical case, to drive the MBHs to coalescence in 
less than a Hubble time \cite{MMS07}.

Nevertheless, the spherical approximation is not only a {\it worst case
scenario} from the point of view of the loss cone dynamics, but it is also
highly unrealistic in the sense that no merger remnant will ever exhibit 
such a high degree of symmetry. It is known that some significant (transient) 
triaxiality results almost inevitably from a galaxy  merger \cite{Moore04} --- 
as well as some net rotation, details depending on stellar and gas content, orbital 
parameters of the merger, etc. We have recently modelled the galactic nucleus
remnant with rotating King models; in some cases, the rotation was high enough
that the model was unstable to the formation of a rotating, triaxial bar-like
feature. In such setting, the evolution of the MBH binary did not stall at $a \approx
a_h$, but did inspiral down to coalescence on timescales well below a Hubble
time \cite{BMSB06,BP08}. 

Notwithstanding the improvement of the latter models as compared to spherical
ones, we have decided to take the next step further and
set up a number of galaxy mergers \cite{PB08}. We let two galactic nuclei to 
merge on nearly parabolic orbits, since this seems to be 
a typical configuration for mergers as is observed in state-of-the-art
cosmological simulations \cite{KhoBur06}. Each galactic nucleus is represented
by a spherically symmetric model with a power law density cusp of logarithmic 
slope $\gamma$ at the center, and a break radius which we set to
unity in model units \cite{De93,Tr94};
incidentally, this also permits us to study the mass deficits resulting from the
ejection of stars by the binary. A massive point particle represents the MBH
which, at the beginning, is placed in the center of each nucleus with zero 
velocity with respect to each nucleus. The total mass in stars of each galactic
nucleus is unity; we adopt units where $G=1$. 
In our sample, we have currently some sixty simulations in which we vary
essentially four parameters: the total mass of 
the MBH binary $M_{12}=M_1+M_2$ in units of the total stellar mass in each merger
progenitor, its mass ratio $q=M_1/M_2<1$, the central slope $\gamma$ of the
stellar density of each galaxy and the total number $N=N_1+N_2$ of stars in the 
two galaxies. We have generated a sample of simulations for the following
parameter range: $5 \times 10^{-3} \leq M_{12} \leq 6 \times 10^{-2}$, $0.2
\leq q \leq 1$, $0.5 < \gamma \leq 1.2$ and $64K \leq N \leq 256K$.

\begin{figure}[h]
\vspace{-3.0pc}
\begin{minipage}{18pc}
\includegraphics[width=18pc]{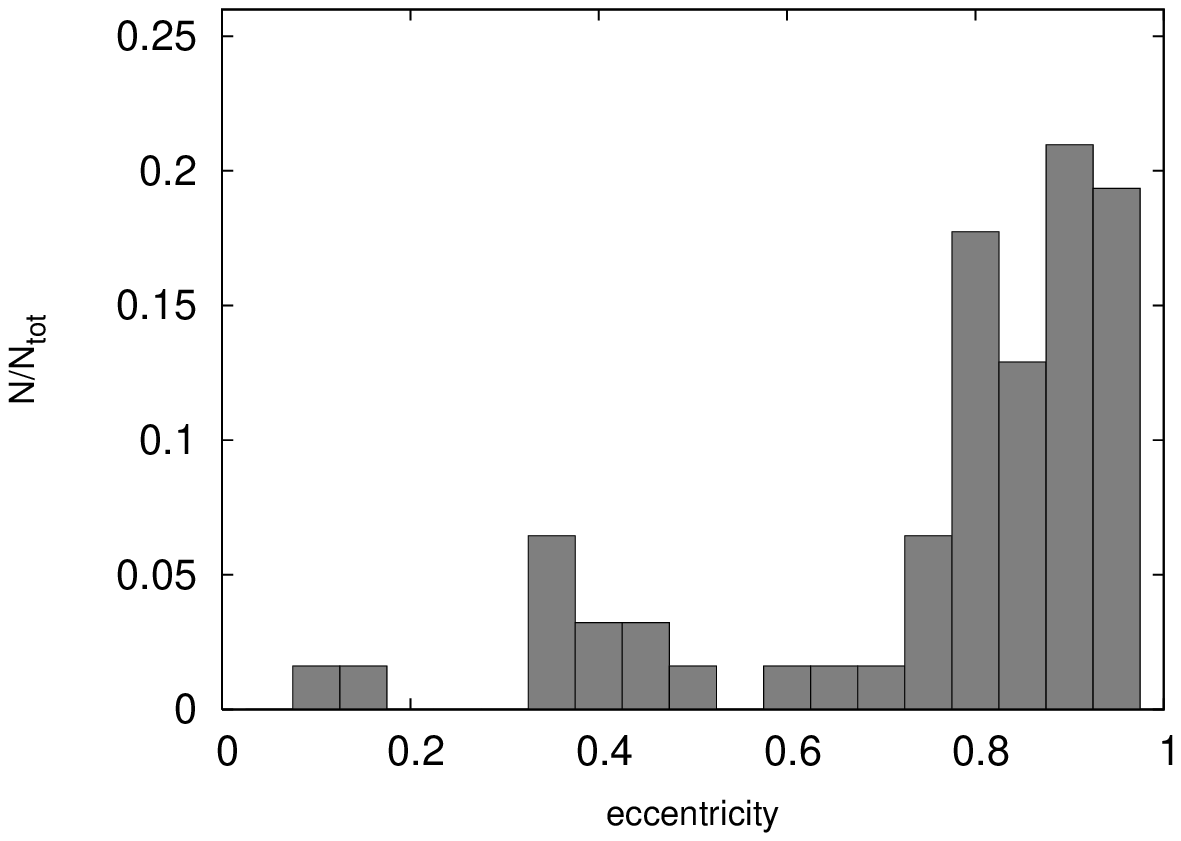}
\caption{\label{label}Histogram representing the eccentricity distribution
 with which the MBH binaries settle when they form a bound
pair. This value is an average value taken over a few $N$-body time units,
after the binary becomes "hard".}
\end{minipage}
\hspace{1.0pc}
\begin{minipage}[s]{18pc}
\vspace{+2.5pc}
\includegraphics[width=18pc]{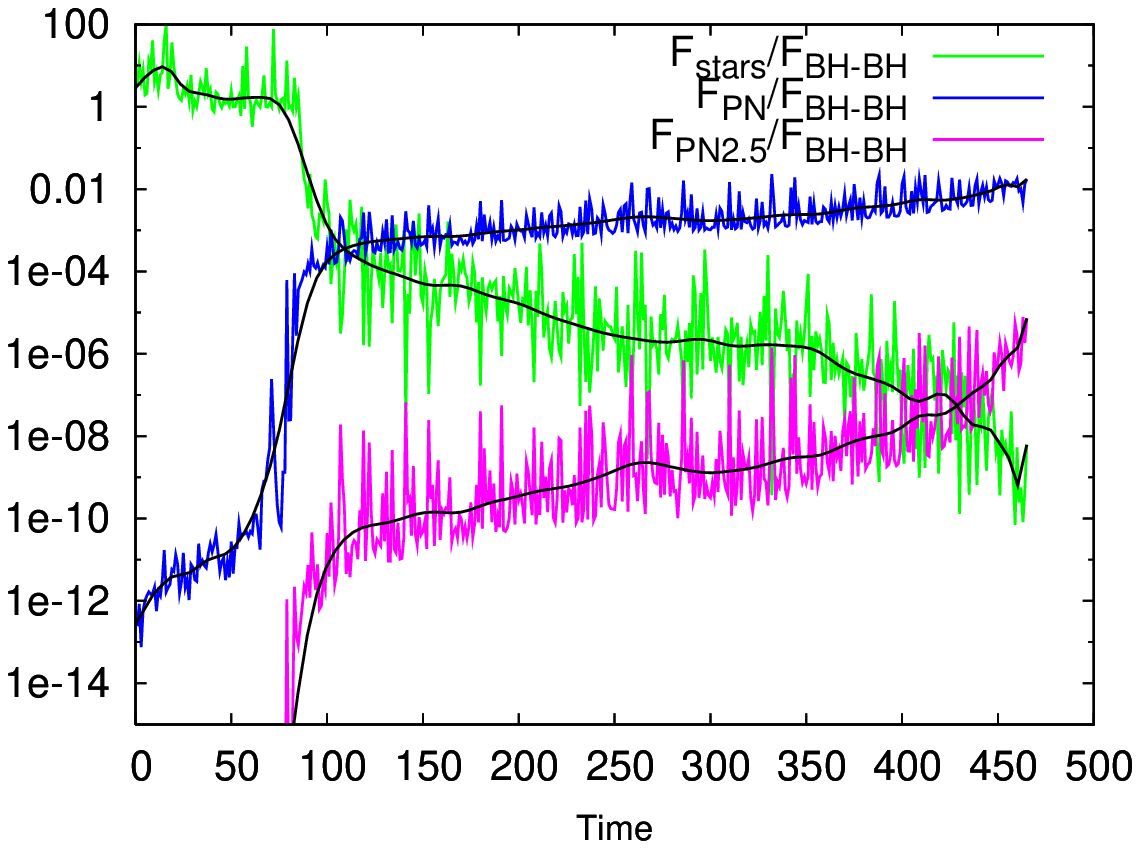}
\caption{\label{label}Fractional value of the perturbing accelerations, 
normalized by the Newtonian acceleration between the MBHs. From: (a) the stars (green), 
(b) the total 1PN+2PN+2.5PN terms (blue), (c) the 2.5PN term only (magenta). 
The perturbing forces are always orders of magnitude weaker than the dominant 
Keplerian force.}
\end{minipage} 
\end{figure}

After the two nuclei have merged, the MBHs sink to the center of the remnant
where they quickly form a bound pair. In Fig. 1, we show the statistics for the
average eccentricity with which the MBHs form a bound pair in our sample.
There is a broad range of possible eccentricities, but with a clear bias
towards the formation of binaries with very high eccentricity $e \geq 0.8$. 
This tendency for the formation of highly eccentric binaries results naturally 
from the (cosmologically motivated) initial conditions with which we set up
the merger: the nuclei fall into each other on near-parabolic orbits;
therefore the MBHs, being much heavier than the stars, also approach each
other on near-parabolic orbits despite the 
perturbations from the surrounding stellar field. In our previous work, we
have modelled the galactic nucleus of a merger remnant with rotating King 
models; in such models, the MBH binaries also
tend to become bound with very high eccentricity \cite{BMSB06,BP08}. In the latter
work, it has been shown that the hardening rate due to the superelastic
scattering of field stars is essentially independent of the eccentricity. It
is well known that the emission of GWs has a very steep dependence
on the pericenter distance and thus a highly eccentric binary inspirals on a much
shorter time scale \cite{P64}.

\section{Orbital evolution of the massive black hole binary}
We have implemented the relativistic effects to the MBH binary
only, by using the Post-Newtonian (hereafter PN) equations of
motion written in the center of mass frame including all terms up 
to 2.5PN order:
\begin{equation}
\frac{d {\bf v}}{dt} = -\frac{G M_{12}}{r^2} \left[(1+\cal A) {\bf n} +
  \cal B {\bf v} \right] + {\cal O} (1/c^6),
\end{equation}
where ${\bf n}={\bf r}/r$, the coefficients ${\cal A}$ and ${\cal B}$ are
complicated expressions of the binary's relative separation and velocity
\cite{B06,BI03}. The Post-Newtonian approximation is a power series 
expansion in $1/c$: the $0^{th}$ order term corresponds to the dominant Newtonian
acceleration; the even order 1PN and 2PN terms are conservative and
proportional to $1/c^2$ and $1/c^4$ respectively, and they are responsible for
the precession of the pericenter; finally, the lowest order dissipative 2.5PN
term which is proportional to $1/c^5$ causes the loss of orbital energy and of
angular momentum due to radiation reaction. We treat the MBHs as point
particles and thus we neglect 
any spin-orbit or spin-spin coupling. The runs were carried --- either with the
direct-summation NBODY4 \cite{Aar99} or $\varphi$-GRAPE \cite{Harfst07} codes --- 
on the high-performance GRAPE-6A cluster at the Astronomisches Rechen-Institut 
(Heidelberg). The MBH
pair motion, once it becomes a hard binary, is integrated with the $4^{th}$ 
order Hermite scheme, after performing
a transformation to KS regularized variables \cite{Aar99,Aar03}. The
PN corrections are treated as perturbations to the dominant
Newtonian acceleration in exactly the same manner as the perturbations from the field
of stars. In Fig. 2, we show that both type of perturbations are, up to the
very late stages of the relativistic inspiral, always much weaker than the dominant
Newtonian acceleration. Therefore we are safe in adopting the linear approximation
when adding both contributions to obtain the total perturbation. Although the 
1PN and 2PN terms are conservative and thus do not drive directly the relativistic 
inspiral, their strength is still, for most of the time, several orders of
magnitude higher than that of the dissipative 2.5PN term. As a result, these
conservative terms should be included in the calculation at all times in order
to get an accurate evolution for the orbital elements during the whole
inspiral and until they reach the LISA band \cite{BP08,BPBS08}.

\begin{figure}
\begin{center}
\includegraphics{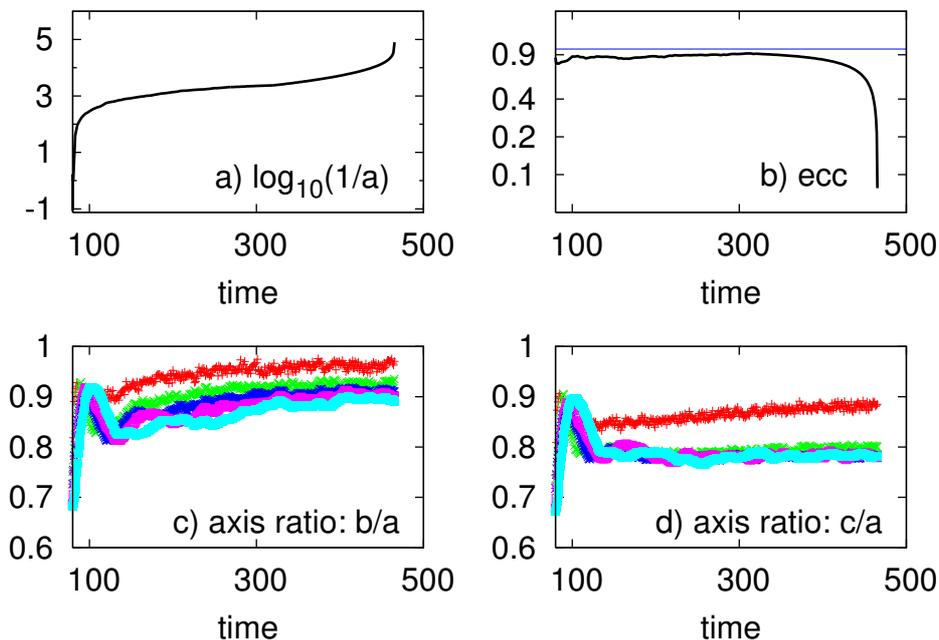}
\end{center}
\caption{\label{label} (a) Evolution of the inverse semimajor axis; (b)
  evolution of the eccentricity; (c) the evolution of the intermediate axis ratio
  $b/a$; (d) the evolution of the
 minor axis ratio $c/a$. The different colors represent five different radii:
 $r=1$ (red), $r=3$ (green), $r=5$ (dark blue), $r=7$ (magenta), and $r=9$ 
 (light blue).}
\end{figure}

We next describe a run that is typical of our sample. In this fiducial run, we
set the black hole masses $M_1$ and $M_2$ of the two MBH particles to be $1\%$
and $2.5\%$ ($q=0.4$) of the total mass in stars. We employ a total of $N=128K$ particles to
represent the total number of equal mass stars (half on each nucleus). Both
nuclei have central densities with logarithmic slope $\gamma=1$. Since LISA
will be sensitive to the inspiral of MBH binaries of total mass 
$M_{12} \leq 10^7 M_{\odot}$, we have chosen to scale our model in such a way
that the lighter MBH has $M_1=10^6 M_\odot$ in physical units. As a result, one
spatial {\it N}-body unit corresponds to $5$ pc and one {\it N}-body time unit corresponds
to $\approx 1.67 \times 10^4$ yr. The resulting speed of light in our model units is
$c=1022$. We place initially the centers of the two nuclei at a separation of $20$
model units, i.e. $100$ pc. The half-mass radius of each nucleus is,
for the fiducial run with $\gamma=1$, $r_{1/2} \approx 2.41$ in model units, or
some $12$ pc; this means that the nuclei are well separated in their initial
configuration. The total mass in stars, initially enclosed within the inner $12$ pc of
each nucleus, is $5 \times 10^7 M_\odot$.

After roughly $80$ {\it N}-body time units, the MBHs
have reached the center of the remnant and form a bound pair. At a first
stage, they eject a large amount of mass ($M_{ej} \sim M_{12}$) in stars and
its semimajor axis has shrunk by a factor of $\sim 50$ by time $t=100$. Subsequently,
the hardening rate slows down considerably but does never stall.  

The binary forms with an eccentricity close to $0.9$, which is typical of our
simulation sample, and grows slowly over time before it eventually starts to 
circularize due to radiation reaction. In Fig. 3a, we can see that the MBHs 
coalesce after roughly $450$ {\it N}-body
units, or some $7.5$ Myr. We should provide a word of caution,
though, since the inspiral time scale is very sensitive to the eccentricity evolution 
as a function of the binary's semimajor axis \cite{P64}. The study of the dependence of
the evolution of the eccentricity as a function of the initial conditions, and
especially as a function of the total number $N$ of stars (which, for a fixed total mass, 
determine the mass ratio between the MBHs and the stars) is the subject of  
ongoing work \cite{PB08}.

In panels $3c$ and $3d$, we can observe the evolution of the axis ratio for the mass
distribution of the merger remnant at several different radii. The radius of
influence $r_h$ of the binary --- radius enclosing $2 M_{12}$ of the stellar mass
--- is in {\it N}-body (physical) units 0.2 (5 pc). The overall
shape of the cluster is markedly triaxial throughout the simulation; at the
inner radius $r=1$ represented in the figure, the triaxiality parameter 
$T \equiv (a^2-b^2)/(a^2-c^2)$
decreases faster but is still $\approx 0.1-0.25$ by the time the binary decouples
from the stellar environment at time $t \approx 410-420$ (6.9 Myr).

%\begin{figure}[h]
%%\vspace{-5.0pc}
%\begin{center}
%\includegraphics[width=22pc]{rho.eps}\hspace{1pc}%
%\begin{minipage}[b]{22pc}\caption{\label{label}The scouring effect on the
%central density cusp of the merger remnant as the MBH binary semimajor shrinks 
%through the ejection of stars. The mass loss extends to regions well outside
%of $r_h$, as expected if the loss cone is kept full by a centrophilic family
%of orbits associated with the triaxial stellar potential. The arrow signals
%the binary's radius of influence. Times are: $85$ (black), $90$ (green), $100$
%(dark blue), $150$ (magenta), $320$ (light blue), and $450$ (orange).}
%\end{minipage}
%\end{center}
%\end{figure} 

\begin{figure}[h]
\includegraphics[width=22pc]{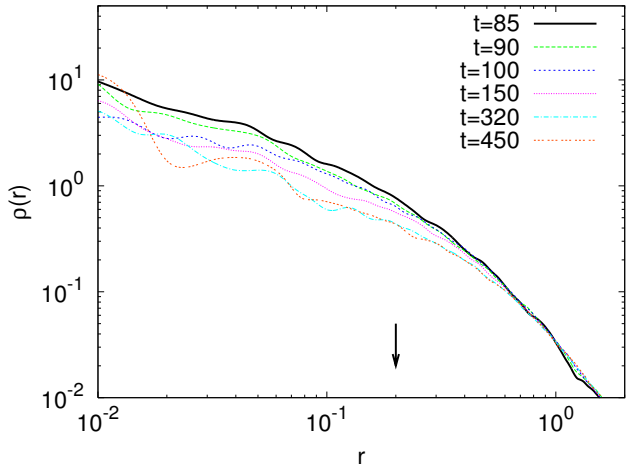}\hspace{1pc}%
\begin{minipage}[b]{14pc}
\caption{\label{label} The scouring effect on the
central density cusp of the merger remnant as the MBH binary semimajor shrinks 
through the ejection of stars. The mass loss extends to regions well outside
of $r_h$, as expected if the loss cone is kept full by a centrophilic family
of orbits associated with the triaxial stellar potential. The arrow signals
the binary's radius of influence.}
\end{minipage}
\end{figure}

As the semimajor axis  of the binary shrinks through the ejection of stars,
the central density decreases progressively. If the triaxial nature of the
potential drives the collisionless refilling of the loss cone, it is expected
that a fair amount of stars supplied to the binary originate from a region
well outside its influence radius $r_h$ \cite{MPoon04}. On the contrary,
in the case of an empty loss cone in a spherical symmetric nucleus, the
scouring effects on the central density cusp would be limited roughly to the
region interior to $r_h$ \cite{M06}. In our model, the influence radius
is initially $r_h \sim 0.18$ and evolves only very slowly to $r_h \sim 0.22$
at the end. From the inspection of Fig. 4, we can immediately see that the
density cusp is being depleted out to radii of $\sim$ few$\times r_h$, and 
this is a generic result in our simulation sample \cite{PB08}. Furthermore, we
measure the amount of mass ejected by the binary, by $t=320$, from the region
interior to $3 r_h$ to be $\approx 2 M_{12}$; in a spherical nucleus, we
expect a much smaller value $M_{ej} \approx 0.5 M_{12}$ \cite{M06}. Note that 
this high rate of mass ejection may help to explain the existence of large
mass deficits in some bright elliptical galaxies.

We can also observe that towards the end of the run, when the MBH pair
has almost completely decoupled from the galactic nucleus ($t \sim 410$) and is
evolving mainly in isolation under the effect of radiation reaction, 
the density cusp does not decay anymore. In fact, the cusp is starting to 
grow again, though at a much slower rate than that with which it decayed
before, as expected in the case of a single MBH. 

Can we extrapolate these results to real galaxies? The answer to this question 
depends on the extent to which the evolution of the orbital elements
(hardening rate and eccentricity evolution) converges for the range of
particle particle number used in our simulations. Our preliminary results 
indicate that the hardening rate does indeed converge once we reach $N \approx
0.125-0.25 \times 10^6$. This is presumably due to the fact that the
repopulation of the loss cone orbits happens in a regime that is essentially
collisionless as a result of the nonlinear dynamics of a centrophilic
family of orbits supported by the triaxial remnant nucleus.

\begin{figure}[h]
\begin{minipage}{18pc}
\includegraphics[width=18pc]{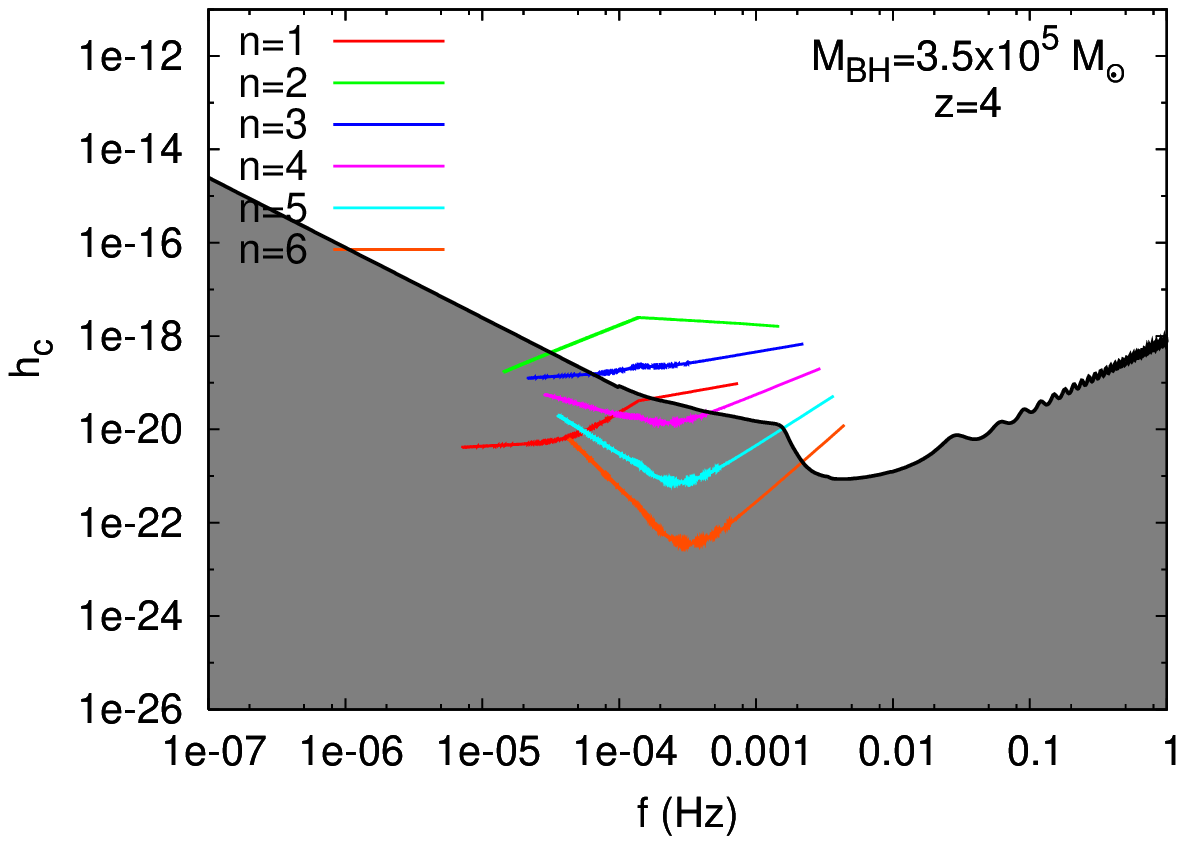}
\caption{\label{label} Locus of the MBH binary, for the first six harmonics,
  in the LISA sensitivity band during their final inspiral and coalescence. 
  The dimensionless strain $h_c$ is plotted against the observed frequency.
  The total mass of the binary is $M_{12}=3.5 \times 10^5 M_\odot$.}
\end{minipage}
\hspace{1.0pc}
%\vspace{-2.0pc}
\begin{minipage}[s]{18pc}
\vspace{-4.0pc}
\includegraphics[width=18pc]{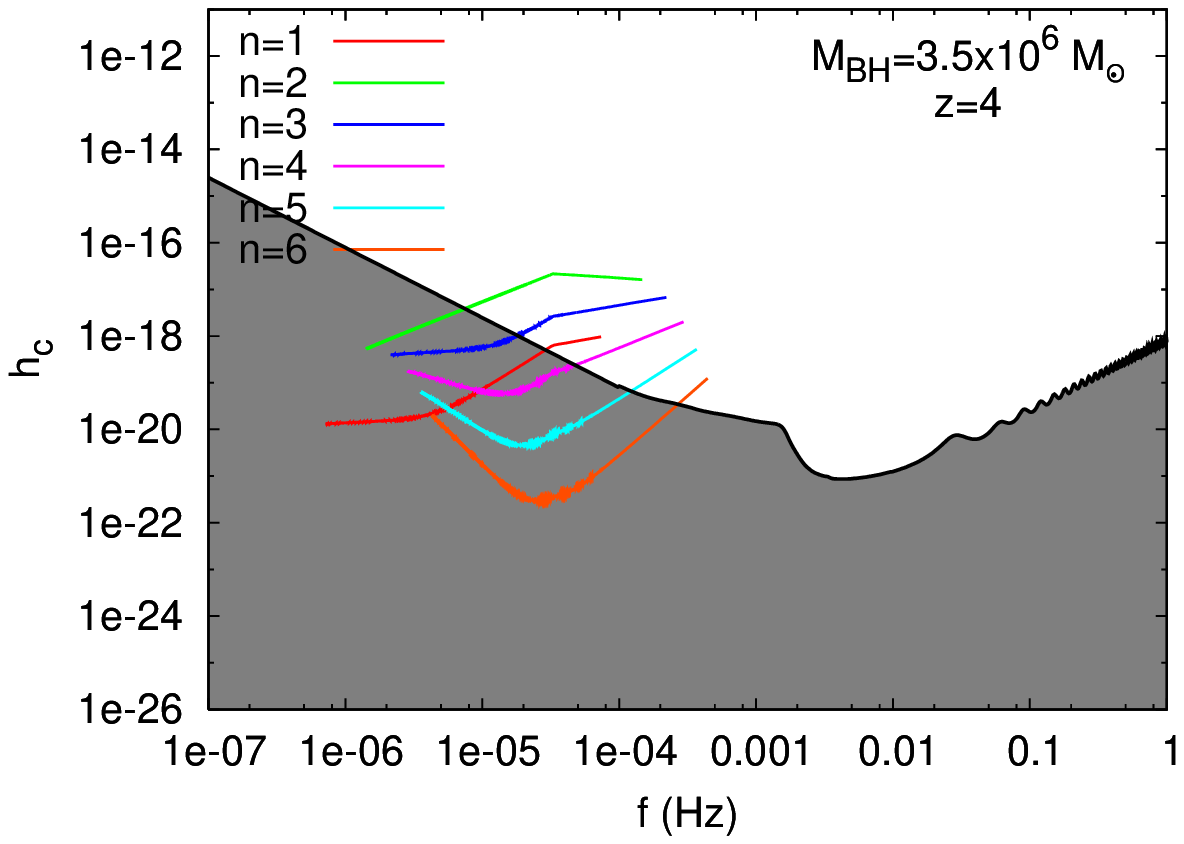}
\caption{\label{label} The same as Fig. 5, but for $M_{12}=3.5 \times 10^6 M_\odot$. }
\end{minipage} 
\end{figure}

\section{Gravitational Wave Signal}
The inspiral and merger of MBHs are expected to be one of the brightest
sources of GWs to be detected by LISA \cite{Da97}. In most of
the literature, the strain amplitude of the GWs is estimated under the
assumption of circular orbits. This has been motivated by the idea that MBH
binaries should have been completely circularized by the time they enter the
LISA band at $f \sim 10^{-4}$ Hz \cite{Sesana05}. In our runs, the
binaries form typically with very large eccentricities and usually reach
small separations (e.g. $100$ $R_{Schw}$) with non-negligible eccentricty. 
Depending on the redshift of the source, this implies that they may enter
the LISA band with significant power distributed among the higher harmonics
$f_n = n f_{orb}/(1+z)$, with $n \geq 2$ (note that $n=2$ for a circular
orbit) \cite{PM63}. In Figs. 5 and 6, we plot the characteristic strain 
amplitudes $h_c$ for the our fiducial run when the model is scaled in such a way
that the total mass of the BH binary is $M_{12} = 3.5 \times 10^5 (3.5 \times10^6)
M_\odot$ and if the redshift of the source were $z=4$. We conclude that 
several higher harmonics stand above the LISA sensitivity curve. We suggest 
that MBH binary orbits with non-vanishing eccentricity may be seriously
considered for the LISA data analysis.
 
\section{Conclusions}

We present preliminary results from a set of {\it N}-body models of merging galactic 
nuclei. The generic outcome of these mergers is the formation of a triaxial
remnant. The MBHs present in these models often form highly eccentric
binaries and, driven by the large pool of stars in centrophilic orbits
associated to the triaxial potential, do not stall at a separation $a \approx
a_h$. Those binaries that get formed with the highest eccentricity, $e \geq
0.9$, can decay from initial separations of $\sim 100$ pc
down to gravitational-wave-induced coalescence within $\sim 10^7$ yr. These
binaries may reach the LISA band with non-negligible eccentricity and thus it
may be important to consider finite-eccentricity effects in the data analysis.
The amount of stellar mass ejected by the binary throughout the inspiral is higher
than estimates made for spherical nuclei --- this may help explaining the mass
deficits observed in some of the brightest elliptical galaxies.

{\small
\ack{This work is supported by DLR (Deutsches Zentrum f\"{u}r Luft- und
  Raumfahrt), the Volkswagen Foundation (GRACE Ref. I/80 041-043) and
SFB 439 of DFG. MP acknowledges the Symposium organizers for financial support
  to attend the conference.}

\section*{References}

%\begin{thereferences}

%\end{thereferences}
}


\begin{thebibliography}{10}   

% 10 means I can use up to 99 refs; 100 would mean up to 999; and so on ....


\bibitem{FF05} Ferrarese, L. and Ford, H., {\it Space
    Sci. Rev.}, {\bf 116}, 523-624

\bibitem{Ma98} Magorrian, J. {\it et al.}, {\it
    Astron. Journ.}, {\bf 115}, 2285-305

\bibitem{FM00} Ferrarese, L. and Merritt, D., {\it
  ApJL}, {\bf 539}, L9-12

\bibitem{Ge00} Gebhardt, K. {\it et al.},  {\it
  ApJL}, {\bf 539}, L13-16

\bibitem{BBR80} Begelman M.C., Blandford R.D. and Rees
M.J., 1980, {\it Nature}, {\bf 287}, 307-9

\bibitem{MRev06} Merritt, D., 2006, {\it Rep. Prog. Phys.}, {\bf 69}, 2513-79

\bibitem{AN05} Armitage, P. and Natarajan, P.,
  2005, {\it ApJ}, {\bf 634}, 921-27

\bibitem{Mayer07} Mayer, L., Kazantzidis, S., Madau, P., Colpi, M., Quinn, T.
and Wadsley, J.\ 2007, {\it Science}, {\bf 316}, 1874

\bibitem{PA08} Perets, H.B. and Alexander. T.,
  2008, {\it ApJ}, {\bf 677}, 146-59

\bibitem{MPoon04} Merritt D. and Poon M.Y., 2004, {\it ApJ}, {\bf 606}, 788-98

\bibitem{BMSB06}Berczik, P., Merritt, D., Spurzem and R.,
  Bischoff, H.-P., 2006, {\it ApJ}, {\bf 642}, L21-4

\bibitem{BP08}
Berentzen, I., Preto, M., Merritt, D., Berczik, P. and Spurzem, R., 2008 {\em submitted to ApJ}

\bibitem{Yu02} Yu, Q., 2002, {\it MNRAS}, {\bf 331}, 935-58

\bibitem{MM03}
Milosavljevi\'c M. and Merritt, D.\ 2003, {\it ApJ}, {\bf 596}, 860-78

\bibitem{Hem02} Hemsendorf M., Sigurdsson S.,
Spurzem R., 2002, {\it ApJ}, {\bf 581}, 1256-70

\bibitem{MF04} Makino, J.~\& Funato, Y., 2004, {\it ApJ}, {\bf 602}, 93-102

\bibitem{BMS05}
Berczik, P., Merritt, D. and Spurzem, R., 2005, {\it ApJ}, {\bf 633}, 680-7

\bibitem{MMS07} Merritt D., Mikkola S and Szell A., 2007,
{\it ApJ}, {\bf 671}, 53-72

\bibitem{CK78} Cohn, H. and Kulsrud, R., 1978, {\it
    ApJ}, {\bf 226}, 1087-108

\bibitem{Sesana07} Sesana, A., Volonteri, M. and Haardt, F.,
  2007, {\it MNRAS}, {\bf 377}, 1711-16

\bibitem{Moore04} Moore B, Kazantzidis S., Diemand J. and
  Stadel J., 2004, {\it MNRAS}, {\bf 354}, 522-28

\bibitem{PB08}
Preto, M., Berentzen, I., Berczik, P., Merritt, D. and Spurzem, R., 2008 {\em to
  be submitted}

\bibitem{KhoBur06} Khochfar, S.
  and Burkert, A., 2006, {\it Astron. Astrop.}, {\bf 445}, 403-12

\bibitem{De93} Dehnen, W., 1993, {\it MNRAS}, {\bf 265}, 250-6

\bibitem{Tr94} Tremaine, S., {\it et al.}, {\it
    Astron. Journ.}, {\bf 107}, 634-44

\bibitem{P64}Peters, P.C., 1964, {\it Phys. Rev.} B, {\bf 136}, 1224-32

\bibitem{B06}Blanchet, L., 2006, {\it Living Rev. Relativity}, {\bf 9}, 4. \footnote{URL (cited on June 2007):http://www.livingreviews.org/lrr-2006-4}

\bibitem{BI03} Blanchet, L. and Iyer, B.R., 2003, 
{\it Clas. Quant. Grav.}, {\bf 20}, 755-76

\bibitem{Aar99}Aarseth, S.J., 1999, {\it Publ. Astron. Soc. Pac.}, {\bf 111}, 1333-46

\bibitem{Harfst07} Harfst, S., Gualandris, A., Merritt, D., Spurzem, R.,
  Portegies Zwart, S. and Berczik, P. 2007, {\it New Astron.}, {\bf 12}, 357-77

\bibitem{Aar03}Aarseth, S.J., 2003, Gravitational $N$-body
Simulations (Cambridge: Cambridge Univ. Press)

\bibitem{BPBS08}Berentzen, I., Preto, M.,
  Berczik, P., Merritt, D. and Spurzem, R., 2008, {\it Astronomische
  Nachrichten}, {\bf 329}, 904-7

\bibitem{M06} Merritt, D.\ 2006, {\it ApJ}, {\bf 648}, 976-86

\bibitem{Da97} Danzmann, K., 1997, Clas. Quantum Grav., 14, 1399-404

\bibitem{Sesana05}
 Sesana, A., Haardt, F., Madau, P. and Volonteri, M., 2005, {\it ApJ}, {\bf 623}, 23-30

\bibitem{PM63} Peters, P.C. and Mathews, J., 1963, {\it
    Phys. Rev.}, {\bf 131}, 435-40







\end{thebibliography}
\end{document}